# Soliton-Assisted Massive Signal Broadcasting via Exceptional Points

Zhuang Fan[1†], Yukun Huang[1†], Wenchan Dong[1†], Haodong Yang[1], Jiahao Hu[2], Yizheng Chen[1], Hanghang Li[1], Nuo Chen[1], Heng Zhou[2,4], Jing Xu[1,3,5] and Xinliang Zhang[1,3,6]

[1]*School of Optical and Electronic Information and Wuhan National Laboratory for Optoelectronics, Huazhong University of Science and Technology, Wuhan 430074, China*

[2]*Key Lab of Optical Fiber Sensing and Communication Networks, School of Information and Communication Engineering, University of Electronic Science and Technology of China, Chengdu 611731, China*

[3]*Optics Valley Laboratory, Hubei 430074, China*

[4]*Corresponding author: zhouheng@uestc.edu.cn*

[5]*Corresponding author: jing_xu@hust.edu.cn*

[6]*Corresponding author: xlzhang@hust.edu.cn*

†These authors contributed equally

**Abstract**

Chip-scale all-optical signal broadcasting enables data replication from an optical signal to a large number of wavelength channels, playing a critical role in enabling massive-throughput optical communication and computing systems. The underlying process is four-wave mixing between an optical signal and a multi-wavelength pump source via optical Kerr nonlinearity. To enhance the generally weak nonlinearity, high-quality (Q) microcavities are commonly used to achieve practical efficiency. However, the ultra-narrow linewidths of high Q cavities prohibit achieving massive throughput broadcasting due to Fourier reciprocity. Here, we overcome this challenge by harnessing a parity-time symmetric coupled-cavity system that supports equally spaced exceptional points in the frequency domain. This design seamlessly integrates generation of dissipative Kerr soliton comb source and all-optical signal broadcasting into a unified nonlinear process. As a result, we realize soliton-assisted intracavity massive signal broadcasting with a channel count exceeding 100 over 200 nm wavelength range, resulting in Terabit-per-second aggregated rates. This throughput surpasses the intrinsic microcavity linewidth constraint (~200 MHz) by over three orders of magnitude. We further demonstrate the utility of this approach through an optical convolutional accelerator, highlighting its potential to enable transformative capabilities in photonic computing. Our work establishes a new paradigm for chip-scale photonic processing devices based on non-Hermitian optical design.

**Introduction**

The utilization of wavelength resources has been a key driver behind the explosive growth in capacity of modern optical communication and computing systems[1–3]. As an essential way to manipulate optical information in the wavelength domain, all-optical signal broadcasting performs data replication of an optical signal to a large number of wavelength channels via optical nonlinearity[4–10], unlocking diverse applications including telecommunication band expansion[4,11], channelized radio frequency receiver[12–15], and convolutional neural network acceleration[1,2,7]. Featuring femtosecond-scale response times, all-optical signal broadcasting holds the potential to eliminate the need for costly and power-intensive analog/digital to digital/analog converters (ADCs/DACs) as well as O-E-O (optical-electrical-optical) conversions[11,16–20]. To enable all-optical signal broadcasting, the input signal is mixed via four-wave mixing (FWM) with a multi-wavelength pump laser. The number and coherence of the pump lines critically determine the throughput and noise of the broadcast channels[5,7,9].

Dissipative Kerr soliton (DKS) comb, a natural consequence of self-organized soliton pulse trains in the temporal domain, serves as an ideal pump source due to its outstanding coherence and broadband feature[21–27]. Although the generation of DKS is also based on FWM, current chip-scale broadcast systems rely on the sequential use of resonators and external nonlinear waveguides for DKS generation and broadcasting, respectively[5,7,8]. In this case, the output power of DKS from microcavities is low, typically up to only a few percent of the pump power, due to that fact that the soliton pulses experience a short interaction time with the pump field as well as the large red pump detuning in the mode locking mechanism. Enhanced conversion efficiency may be achieved but at the expense of delicate dispersion engineering as well as complicated tuning procedure[28–33]. Such a design necessitates external optical amplification to boost the power of DKS combs, which not only increases system complexity and energy consumption, but also drastically reduces the number of usable comb lines owing to the limited gain bandwidth of commercially available optical amplifiers. As a result, the number of broadcast channels in current chip-scale broadcasting remains rather limited (up to 28)[4–9]. A possible path to go far beyond this limitation is to integrate comb generation and nonlinear broadcasting within a single resonant device. In this case, the entire broadband intracavity DKS comb, which exhibits significantly high-power levels from resonant enhancement, can be collectively and efficiently utilized as a whole. However, this vision runs into a profound physical contradiction originated from Fourier reciprocity. That is, the high-Q cavities needed for DKS formation demand extreme spectral purity, corresponding to a low information entropy system, while broadband signal processing requires high-entropy, lower-Q environments supporting high-speed dynamics. This information entropy mismatch has rendered intracavity signal broadcasting nearly prohibitive.

Recently, exceptional points (EPs)[34–39]—spectral singularities accessible through parameter variations in open systems—have attracted considerable attention for their ability to induce unconventional phenomena in physical systems, such as chiral optical responses[34,40–42]. In particular, the unique feature of EPs in parity-time (PT) symmetric systems[43–47] indicates an effective way to bridge the gap between low and high entropy system by counter intuitively utilizing the property of losses[47]. Here, we overcome this limitation by exploiting equally spaced EPs in the frequency domain in a PT symmetric coupled-cavity system, which successfully merges soliton generation and signal broadcasting into a unified nonlinear process. With the assistance of the generation of a robust DKS with perfect soliton crystal state, we achieve intracavity broadband broadcasting over 200 nm, given a channel count number of over 100. This offers an aggregated processing capacity exceeding Terabit-per-second rates, surpassing the intrinsic cavity linewidth (~200 MHz) by more than three orders of magnitude. As a concrete application in convolution accelerator, we further experimentally validate its capabilities in image feature extraction. This all-optical signal broadcasting technology, which synthesizes low- and high-entropy systems via non-Hermitian EPs, establishes a new paradigm for high-efficiency, large-throughput, chip-scale photonic processing.

**Results**

**Merging low and high entropy system via exceptional points**

Merging low and high entropy system, i.e. achieving integrated high- and low-Q resonances for optical broadcasting in a single structure, faces two major challenges. First, while microcavities supporting simultaneous high- and low-Q resonances are readily available in a variety of resonators such as multimode or deformed cavities[48,49], the Q-factor contrast typically stems from loss mechanisms associated with distinct spatial mode profiles. This disparity results in a lack of shared real-space overlap, which is critical for efficient FWM between the high- and low-Q modes. Second, enabling high-capacity signal broadcasting via interaction with a DKS requires the high- and low-Q resonances to be periodically distributed in the frequency domain. Such a spectral arrangement entails an alternating pattern of high- and low-Q modes, posing a significant design constraint. We elaborate in the following that our approach successfully addresses both issues by extending the strategy from our previous design[47].

As shown in Fig. 1, a main resonator is coupled to an auxiliary resonator (with coupling rate $g$) which further couples to a bus waveguide (with coupling rate $\gamma_c$). The intrinsic decay rates of the main and auxiliary resonator are $\gamma_1$ and $\gamma_2$, respectively. By adding significant coupling decay to the auxiliary resonator via $\gamma_c$, i.e., $\gamma_c \gg \gamma_1 \approx \gamma_2$, a passive PT symmetric system can be formed by noting the fact that the subsystem formed by the bus waveguide and the auxiliary resonator acts effectively as a high loss sub-system, in contrast to the main resonator which acts as the low loss counterpart[50]. According to temporal coupled-mode theory (TCMT)[43], the non-Hermitian Hamiltonian of the system reads $\widehat{H} = \begin{bmatrix} \omega_0 - i\frac{\gamma_1}{2} & g \\ g & \omega_0 - i\frac{\gamma_2+\gamma_c}{2} \end{bmatrix}$ when resonances of the two cavities align at $\omega_0$. The eigen-frequencies of the PT-symmetric system can be expressed as $\omega_\pm \approx \omega_0 - i\gamma_c/4 \pm \sqrt{16g^2 - \gamma_c^2}/4$. In the vicinity of the EP, i.e., $\gamma_c \approx 4g$, resonances with super-broadened linewidth that far exceeds the intrinsic linewidth of the resonators, i.e., $\gamma_c$, are synthesized[47]. By setting the radius of the auxiliary cavity as half of that of the main cavity, one in every two resonances of the main cavity undergoes effective broadening via the EP, forming periodic broad resonances, i.e., low-Q modes, for processing high-speed signals. Note that the eigenvector at EP follows $|1,2\rangle = (1, -i)$, where indices 1 and 2 refer to the main and auxiliary resonator, respectively, indicating that in this case the lights resonate in both resonators. This is an important feature that enables effective FWM for broadcasting process, as discussed later. Moreover, our system is tolerant of deviations from the exact EP, a key feature that ensures robust operation of our scheme[47].

While the generation of broad resonances is essential, achieving periodic narrow resonances, i.e., high-Q modes, is equally critical for DKS generation. The remaining half of the resonances, unaffected by the EP, retain their intrinsic high Q-factors and serve as ideal candidates for generating soliton combs. Briefly, the auxiliary resonator can be regarded as an effective waveguide segment at these specific frequencies—a consequence of its anti-resonant condition. Light at these frequencies travels sequentially through two coupling regions, i.e., first from bus to auxiliary resonator and then, from auxiliary to main resonator, leading to a highly favorable situation. That is, these resonances operate at quasi-critical coupling conditions[47], enabling maximized resonant enhancements in the main cavity and maintaining high-Q properties with linewidths of $\sim 2\gamma_1$. Therefore, alternating broad and narrow resonances, i.e., low- and high-Q modes, respectively, with a huge linewidth contrast between the two cases, i.e., $\gamma_c/2\gamma_1$, can be created in our design, ensuring perfect spatial mode overlap of the two contrasting entropy systems for efficient FWM within the main resonator. Field distributions in the narrow and broad resonances are also confirmed by full-wave simulations (given by Fig. S2 in SI-S1, SI is short for supplementary information). More explanations and derivations are given in SI-S1.

The power enhancement spectrum in the main resonator is illustrated in Fig. 1, where the narrow and broad resonances are distinguished by the blue and red color, respectively. The power enhancement is defined as the intracavity power divided by the input power in the bus waveguide. Provided that the effective cavity circumferences are carefully matched, the alternating narrow and broad resonances can extend across the entire power enhancement spectrum. The system operates by injecting a continuous-wave pump and a high-speed signal into the bus waveguide, each aligned with a narrow and a broad resonance, respectively. Such an input, in turn, leads to a scenario where the set of narrow resonances generates a DKS comb, while the set of broad resonances enables high-speed signal broadcasting via cascaded FWM between the high-speed signal and the DKS comb in the main resonator. Such a procedure is illustrated in the inset of Fig. 1, as well as the temporal waveforms shown on top of the main resonator. Finally, both the broadcast data channels and the DKS are routed to the output through the bus waveguide.

**Transmission spectrum and broadcasting bandwidth characterization**

We first characterize the static properties of our device, which is fabricated on a commercially available silicon nitride platform (LIGENTEC). The transmission spectrum of the PT symmetric coupled-cavity system over the 1530-1600 nm wavelength range is shown in Fig. 2(a). For consistency with Fig. 1, the narrow and broad resonances are distinguished by the blue and red color, respectively. The optical microscope image of the device is presented in Fig. 2(b). To align the microcomb spacing with the ITU dense wavelength division multiplexing (DWDM) grid, the free spectral range (FSR) of the main resonator is designed to be 100 GHz. Furthermore, a microheater is integrated above the main resonator to thermally tune and align the resonances of the two cavities. By fitting the transmission spectrum with analytical expressions, device parameters are derived, with 0.6 GHz for both $\gamma_1$ and $\gamma_2$ and 98 GHz and 25 GHz for $\gamma_c$ and $g$, respectively, satisfying quasi-EP condition mentioned earlier (i.e., $\gamma_c = 4g$ for EP) (detailed information is given in SI-S2). Note that in order to ensure sufficiently high coupling coefficients, i.e., large values of $\gamma_c$ and $g$, both the main and auxiliary microresonators are designed in a racetrack geometry. To ensure high intrinsic Q of the resonators, mode crosstalk at the junctions between straight and curved waveguide segments must be minimized, which is achieved by employing Euler curves in the design of the bending sections[51,52].

Figure 2(c) displays a zoomed-in view of the transmission spectrum around 1550 nm. The narrow resonances exhibit a linewidth of approximately 200 MHz, corresponding to a loaded Q-factor of $1\times10^6$ and an estimated propagation loss of the microcavity waveguides of 0.12 dB/cm. The adjacent narrow resonances are spaced 200 GHz apart, which is twice the FSR of the main resonator. The broad resonance in between, as highlighted by the red circle in Fig. 2(c), is hardly observable on the transmission spectrum, due to the highly over-coupled condition induced by the excessively large coupling rate $\gamma_c$. Nevertheless, the power enhancement of the broad resonance is still significant (see theoretical analyses given SI-S3), which contributes to pronounced FWM. In order to characterize the linewidth of the broad resonances, a pump-probe scanning procedure is adopted. That is, a pump light is tuned into the narrow resonance that is used to excite solitons, while the frequency of a probe light is scanned across one of the broad resonances under test. A strong idler is generated at the symmetrically positioned, broad resonance on the opposite side of the pump resonance relative to the probe light. The power of the generated idler light is recorded to calculate the CE, which is defined here by the generated idler power divided by the probe power, both measured on the output FWM spectrum. The measured CE spectrum as a function of the probe detuning is shown in Fig. 2(d) (more linewidth measurement results are given in SI-S7). The linewidth of the broad resonance is defined as the 3 dB bandwidth of the CE spectrum. Figure 2(a) shows the measured linewidths of broad and narrow resonances within the 1530-1600 nm range, as indicated by the red and blue dots, respectively. It can be seen that both broad and narrow resonances exhibit consistent linewidths across the whole measured wavelength range, approximately 15 GHz and 200 MHz respectively, yielding a linewidth contrast ratio of 75:1.

**Dissipative Kerr soliton comb generation and soliton-crystal state formation**

We proceed to experimentally characterize the formation of DKS comb. To achieve anomalous dispersion, the cross-section dimension of the waveguide is designed as 800 nm (height) × 1650 nm (width). The dispersion of the narrow resonances is characterized by the resonance frequency deviation, as depicted in Fig. 3(a), where the integrated dispersion $D_{\text{int}}$ is defined as $D_{\text{int}} = \omega_\mu - \omega_0 - 2D_1\mu$. Here, $\mu$ denotes the relative mode number of narrow resonances with respect to the pumped mode $\omega_0$, and $D_1/2\pi$ corresponds to the FSR at the pump frequency. The experimentally measured $D_{\text{int}}$ values of different mode numbers are represented by dots, while the solid curve illustrates the fitting to the experimental data. From this fit, we extract a second-order dispersion coefficient of $\beta_2 = -99.2$ ps²/km. To generate stable soliton microcombs, an auxiliary laser heating technique is employed[53]. This involves injecting the primary pump light into the bus waveguide while simultaneously introducing a counter-propagating auxiliary laser from the opposite side to stabilize the internal temperature of the main resonator.

By adiabatically tuning the pump wavelength from effective blue- to red-detuned regime, a distinct soliton step at the effective red-detuned regime is observed, as shown in Fig. 3(b). Moreover, deterministic generation of perfect soliton crystals (PSCs) with programmable soliton numbers is achieved. The excitation of PSCs is particularly valuable as they significantly enhance the power per comb line, which is critical for improving the efficiency of signal broadcasting[54–57]. This controllability arises from the interference between the pump light and the backscattered light from the auxiliary laser, which establishes a periodic intra-cavity potential field. The period number of this potential field equals the mode number spacing between the auxiliary light and pump light, and determines the soliton number $N$ of the PSCs[56]. Therefore, by controlling the mode number of the auxiliary light injection, PSCs with desired $N$ can be deterministically generated. Figure 3(c)(i)–(iii) presents the measured optical spectra of the PSCs with soliton numbers of 2, 4, and 6, respectively. The low-power comb teeth observed between the primary soliton lines in Fig. 3(c)(iii) originate from the backscattered light of the modulation instability (MI) state comb induced by the auxiliary light, rather than from defects within the soliton crystal structure. Moreover, the single-soliton state is eliminated in the main resonator when the two cavities are aligned, resulting in a minimum soliton number of $N = 2$. This is a result of significant broadening in half of the narrow resonances of the main resonator, which effectively forms a DKS generation cavity with an FSR twice that of the original. An interesting but natural consequence of this effect is that the soliton number in the main resonator is always an even number.

The soliton number $N$ introduces a trade-off between CE and channel count $M$, i.e., number of broadcast channels. The PSCs with higher soliton numbers exhibit greater per-line power but also larger comb tooth spacing[54,55]. In the context of signal broadcasting, this implies that higher CE can be achieved at the cost of a reduced number of broadcast channels. Specifically, the relationship between CE and soliton number $N$ can be derived as follows (see SI-S5 for the derivation process):

$$\text{CE} = \left(\gamma P_\text{p} \eta L_\text{eff} PE_\text{n} PE_\text{b}\right)^2 N^2, \tag{1}$$

where $\gamma$ represents the nonlinearity coefficient, $P_\text{p}$ the power of the pump light, $\eta$ the conversion efficiency in PSCs generation, $L_\text{eff}$ the effective length of the main ring. $PE_\text{n}$ and $PE_\text{b}$ represent the intra-cavity power enhancement of narrow and broad resonances, respectively. The relationship between $M$ and $N$ can be estimated by:

$$M = \frac{2\text{BW}}{N * \text{FSR}}, \tag{2}$$

where BW represents the frequency range over which signal broadcasting occurs. Note that this formula is applicable for the case of $N \geq 4$. In the case of $N = 2$, each channel is occupied by two broadcast signals generated from distinct nonlinear processes, corresponding to phase-sensitive processes[58,59], resulting the total number of broadcast channels remains identical to that of the $N = 4$ case. Numerical simulation results of the signal broadcasting spectra using different soliton states are presented in Figs. 3(d)(i)–(iii). The input signal and broadcast lights are plotted in red. Figure 3(e) summarizes the relationship of $M$ and CE with respect to $N$, where the dots and solid curves represent simulation and analytical results, respectively. These trends indicate that the 4-soliton crystal state offers an optimal choice, which maximizes the number of broadcast channels while maintaining high conversion efficiency. Therefore, we select the 4-soliton crystal state, i.e., $N = 4$, for subsequent broadcasting experiments. More detailed simulations and explanations are given in SI-S4.

**Massive signal broadcasting up to Terabit-per-second aggregated rates**

The experimental setup used for massive signal broadcasting is shown in Fig. 4(a). To generate the PSCs, pump light from tunable laser TL1 is polarization-controlled, amplified to 26 dBm using an EDFA, routed through a wavelength division multiplexer (WDM), and finally coupled into the waveguide of the broadcasting chip via a lens. The pump wavelength is swept across a narrow resonance near 1557 nm to excite a 4-soliton crystal state. For simplicity, the schematic diagram omits the continuous-wave auxiliary laser at 1554 nm used for thermal stabilization (a similar setup is given in Fig. S15 in SI-S8). A 15 GBaud On-Off Keying (OOK) electrical signal, generated from a bit pattern generator (BPG), is modulated onto an optical carrier wave at 1551 nm from a second tunable laser (TL2) using a Mach–Zehnder modulator (MZM), amplified to 28 dBm by another EDFA, combined with the pump light via the WDM device, and finally injected into the bus waveguide. At the output, a tunable bandpass filter (TBPF) is used for channel selection. The output spectra are measured with an optical spectral analyzer (OSA). The output temporal waveforms and eye diagrams of the broadcast channels are monitored using an oscilloscope (OSC). It is worth mentioning that although the input signal light has high power, its intra-cavity power enhancement is approximately 16 dB lower than that of the pump light. As a result, the intra-cavity power of the signal light remains significantly lower and therefore does not destabilize the soliton comb. The stability of the comb is verified by measuring the RF spectrum, which remains stable with no observable change upon introduction of the signal light (see SI-S8 for RF spectrum measurements).

The output spectrum of signal broadcasting is shown in Fig. 4(b). Blue and red colors are used to distinguish the PSCs comb as well as the original signal and its replicas. It can be seen that the original signal is broadcasted into a large number of equally spaced wavelength channels, with the highest measured CE reaching -38 dB. Over 100 broadcast channels can be clearly observed above the noise floor on the output spectrum, with a total of 62 copies having an output power above -36 dBm (the receiver sensitivity threshold, 10 dB less power shown on the spectrum due to 1:9 power splitting before entering the OSA), including 18 in the S-band, 21 in the C-band, and 24 in the L-band. Due to the limited operational wavelength range of the receiver, only C- and L-band signals are detected. These channels (Ch1–Ch46) are labeled with the order of increasing wavelength, as shown in Fig. 4(c), where Ch16 corresponds to the original signal channel. Elevated noise power near the pump and signal wavelengths —attributed to amplified spontaneous emission (ASE) from EDFAs filtered by a 4.5 nm WDM filter — is also visible. Notably, Ch19, symmetric to the signal wavelength relative to the pump, exhibited significantly lower power. This phenomenon occurs due to the destructive interference between different nonlinear processes at Ch19, which is governed by the specific phase relationship among the lines of the PSC comb (see SI-S4 for detailed simulations and explanations).

The bit error rates (BERs) of broadcast signals are measured for each channel (Ch1 – Ch46). A variable optical attenuator (VOA), an optical pre-amplifier and a BER analyzer (not shown) are used for BER measurements. Results are shown in Fig. 4(d). Solid red dots denote the BERs of the broadcast OOK copies, while red open circles represent the original signal. Ch15, Ch17, and Ch19 exhibit excessively high BER (beyond measurable limits) due to in-band ASE noise and destructive interference. All the remaining channels, a total of 42, display clear eye diagrams (examples for Ch30, Ch40, and Ch46 shown in the inset of Fig. 4(d)), and exhibit BER below $3.8\times10^{-3}$, i.e., the hard-decision forward error correction (HD-FEC) threshold, confirming successful signal replication across 42 channels in the C and L bands. The power penalty for broadcasting is approximately 1.9 dB. More eye diagrams and penalty measurement details are given in SI-S6. By narrowing the WDM bandwidth to filter out the ASE noise in Ch15, and Ch17, and including S-band channels with output power greater than -36 dBm, the total number of usable channels is expected to increase to 62.

A major advantage of signal broadcasting based on the FWM effect is modulation format transparency. Hence, we further experimentally demonstrate the broadcasting of an advanced modulation format involving multi-level phase information, i.e., Quadrature Phase Shift Keying (QPSK) signal, at 15 GBaud. An optical modulation analyzer (Keysight N4392A) is used to analyze the signal quality after broadcasting. Due to the limited operational range of the optical modulation analyzer (1530 – 1568 nm), only Ch1–Ch22 are characterized. Consistent with the OOK results, all channels except Ch15, Ch17, and Ch19 exhibit well-separated constellation diagrams, with Ch10 and Ch20 shown in the sets of Fig. 4(d) (more constellation diagrams are given in SI-S6). The error vector magnitude (EVM) is used to quantify the signal quality. For standard 7% overhead FEC, an EVM below 38% is required for error-free QPSK detection. As indicated by the black solid squares in Fig. 4(d), the highest EVM measured is 34.8% (Ch14), which is still below the detection limit. The black open square shows the EVM of the original signal. In summary, we have demonstrated that our broadcast scheme is compatible with both amplitude and phase data modulation formats. Considering 15 GBaud rate, QPSK data format and a C+L broadcast channel number of 42, the aggregated processing capacity reaches 1.26 Terabit-per-second, exceeding the intrinsic linewidth limit of main cavity (200 MHz) by over three orders of magnitude.

**Matrix convolution accelerator based on massive signal broadcasting**

This demonstrated signal broadcasting scheme offers a significant advantage for high-speed signals that natively or already exist in the optical domain: it dramatically increases their wavelength parallelism. By processing these signals entirely within the optical domain, the scheme avoids electronic bottlenecks and presents a compelling approach for advanced optical computing. This is particularly beneficial for optical convolutional accelerators (OCAs), which significantly enhance computational speed by multiplexing operations in both time and wavelength domains[60,61]. The computational power in such systems is directly proportional to the number of wavelength channels and the signal bit rate. Therefore, the efficient, high-capacity broadcast scheme we demonstrate is a key enabler for advancing OCA performance.

We thus validate the capabilities of our scheme for OCA using an experimental setup shown in Fig. 5(a). A 512×512 image is first binarized and flattened before being encoded onto an optical carrier. As a proof of concept, nine broadcast wavelength channels (Ch2, 4, 6, 8, 10, 12, 14, 16, and 18) are extracted for further operation. A programmable filter (Finisar Waveshaper 1000s) is used to add channel-specific attenuation, thereby encoding a 3×3 convolutional kernel onto the selected wavelength channels. Channel number can be further increased when programmable filter in the L-band is available, where our device currently works only in the C-band. Subsequently, a 2 km single-mode fiber (SMF) is used to provide chromatic dispersion-induced relative delays among the channels with a step size equal to the symbol period. The symbol rate used here is 8 GBaud, corresponding to a symbol period of 125 ps. The symbol rate can be further increased, which is ultimately limited by the linewidth of the broad resonances. Finally, the convolved results are achieved by combining the delayed and weighted optical signals using a high-speed photodetector to execute sum operation[2].

Images processed by different convolution kernels can produce different effects. Here, by reconfiguring the attenuation profile of the programmable filter, four different convolutional kernels are applied, and the corresponding processing results are presented in Fig. 5(b). To evaluate the computational accuracy, the experimental results are compared with theoretical convolution outputs, yielding linear correlation coefficients of 0.927, 0.921, 0.922, and 0.913 for the four kernels, respectively[62]. Thus, this experiment successfully demonstrates accelerated convolutional computation using our broadcasting system, highlighting its application advantages in advancing the parallelism of optical computing.

**Discussions**

Table 1 compares our work with the state-of-art on-chip all-optical signal broadcasting schemes. It can be seen that our solution far surpasses existing schemes in terms of channel count, broadcasting range, and aggregated bit rate, while being extremely concise in structure. While in Table 1 the aggregated bit rate is calculated based on the available channel count in the C+L band in our experiment, by further improving filtering and S-band detection facilities, the total number of usable channels could be increased to 62, leading to an aggregated bit rate of 1.86 Terabit-per-second. By improving the intrinsic Q-factor of the system[63] as well as resorting to material platform with higher nonlinear coefficient, such as III-V materials[9,47], number of channels capable of practical usage can be increased owing to a higher CE. By further broadening the linewidth of low-Q resonances, the single-channel bit-rate can be increased, thereby increasing the aggregated bit rate as well as the computational capacity of optical convolution accelerators. The linewidth of low-Q resonances can be further extended, yet at the cost of balancing the intrinsic Q factor, FSR and complexity of the coupled cavity system. New coupling mechanism and structure design are required to further solve these trade-offs.

In addition, we highlight that the broad resonances created by EPs in our scheme exhibits high tolerance with respect to fabrication derivations from designed coupling coefficients. This is because the linewidth of the broad resonances—a synthesized linewidth—can be maintained near its maximum over a relatively broad range close the EP in the PT-symmetric regime, owing to the merging effect of the two splitting mode peaks[47]. Thanks to this robust feature, we were able to exploit the properties of the EP stably over a 200-nm range to construct the periodic broad resonances, thereby enabling the broadcasting process of over 100 channels, a result that is consistently observed across a large number of our fabricated devices.

Furthermore, our solution offers superior resistance to phase noise accumulation. The accumulation of noise has long posed a significant challenge in signal broadcasting systems, as noise levels increase with the number of channels when the multi-wavelength pumps lack phase correlation. It has been shown that using optical frequency combs with excellent coherence can eliminate phase noise accumulation[58,64–66]. Nevertheless, it is an open issue whether this property exists in this hybrid DKS generation and signal broadcasting design. Thereby, experimental verifications are carried out and we confirm that the phase noise of the broadcast signals does not accumulate with increasing broadcast order (see details in SI-S8).

In summary, we present a powerful design framework that integrates high- and low-entropy systems, enabling the synthesis of DKS generation and high-speed signal broadcasting into a unified process, which is prohibitive in traditional designs. This framework, built upon a PT-symmetric coupled microcavity system, achieves EPs uniformly distributed in the frequency domain alongside alternating high-Q resonances. This non-Hermitian cavity-linewidth manipulation technique essentially enables the realization of an ultra-compact integrated optical signal broadcast system with massive capacity. By demonstrating a 75 times contrast ratio between low-Q and high-Q resonances, we successfully replicate input signals into over 100 wavelength channels across a 200 nm range. Up to 42 measured channels demonstrate performances meeting system-level requirements, achieving an aggregated signal processing rate of 1.26 Terabit-per-second, exceeding the intrinsic microcavity linewidth constraint (~200 MHz) by over three orders of magnitude. Apart from high capacity and simple structure, our device features favorable noise performance, compatibility with advanced modulation formats, and robust operation in the vicinity of EPs—where systems are typically highly sensitive. These combined advantages underscore the overall superiority of our device. Finally, we demonstrate our broadcast scheme for OCA, highlighting its critical role in increasing the wavelength parallelism of native optical high-speed signals. Our work sheds light on solving the universal trade-off between efficiency and speed in signal-processing devices through a non-Hermitian design strategy.

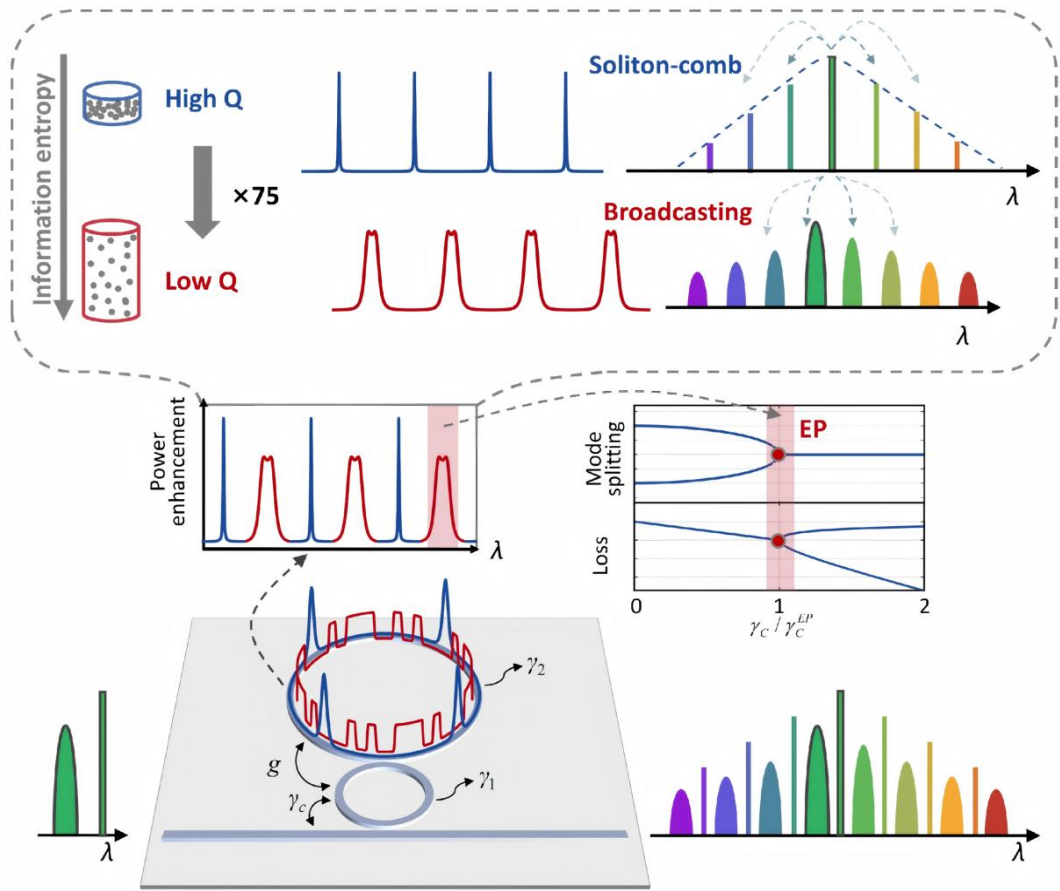

**Fig. 1 Massive signal broadcasting scheme based on equally spaced exceptional points (EPs) in a parity-time symmetric coupled resonator system.** The two coupled microresonators, i.e. the main resonator and the auxiliary resonator, have a circumference ratio of 2:1. When the two resonators are aligned and the coupling parameters are designed to satisfy the EP condition, i.e., $\gamma_c = 4g = \gamma_c^{EP}$, the system gives rise to alternating narrow and broad resonances, i.e. high-Q and low-Q modes, within the main cavity. While the narrow resonances are leveraged to generate stable dissipative Kerr soliton (DKS) frequency combs, the broad resonances, generated via EPs, are exploited to carry and broadcast high-speed data signals. Through efficient nonlinear interaction between the resonantly enhanced DKS comb (excited by launching a pump light into one of the narrow resonances) and the input optical signals (located in one of the broad resonances) within the main resonator, signal replicas are generated in the rest of the broad resonances, achieving massive signal broadcasting.

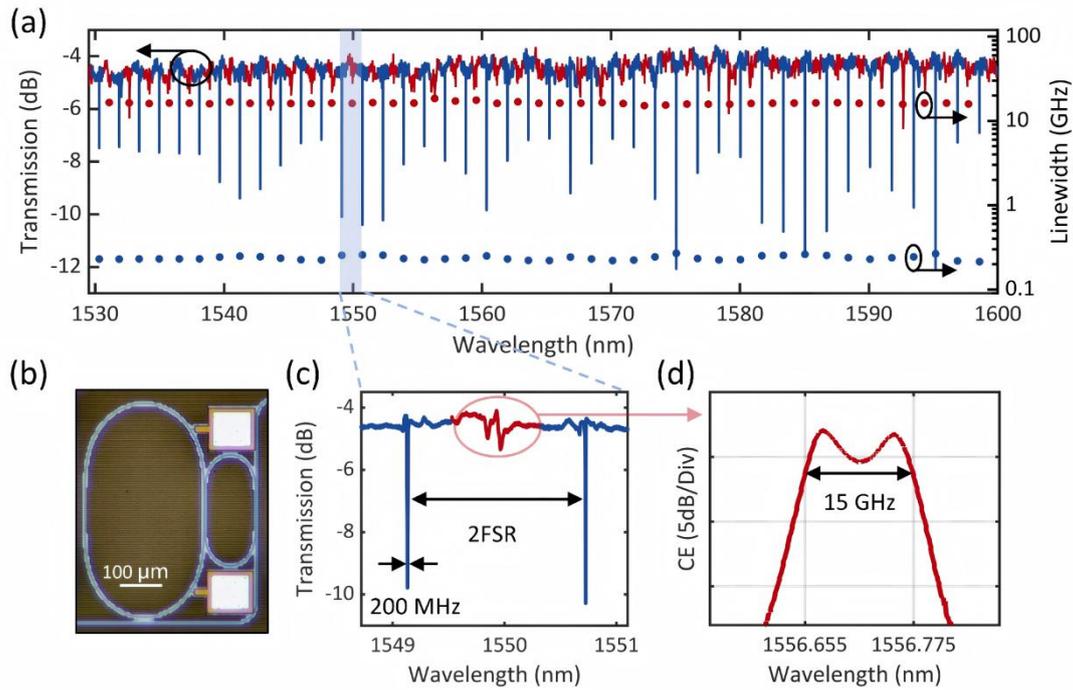

**Fig. 2 Transmission spectrum and broadcasting bandwidth measurements of the PT symmetric coupled resonator system.** (a) The transmission spectrum of the PT symmetric coupled resonator system where the narrow and broad resonances are distinguished by blue and red colors, respectively. (b) The optical microscope image of the device. (c) The zoomed-in transmission spectrum around 1550 nm. (d) Broadcasting bandwidth measurements with a probe scanning method, which is defined as the 3dB bandwidth of the conversion efficiency spectrum. The horizontal axis is the wavelength of the probe light.

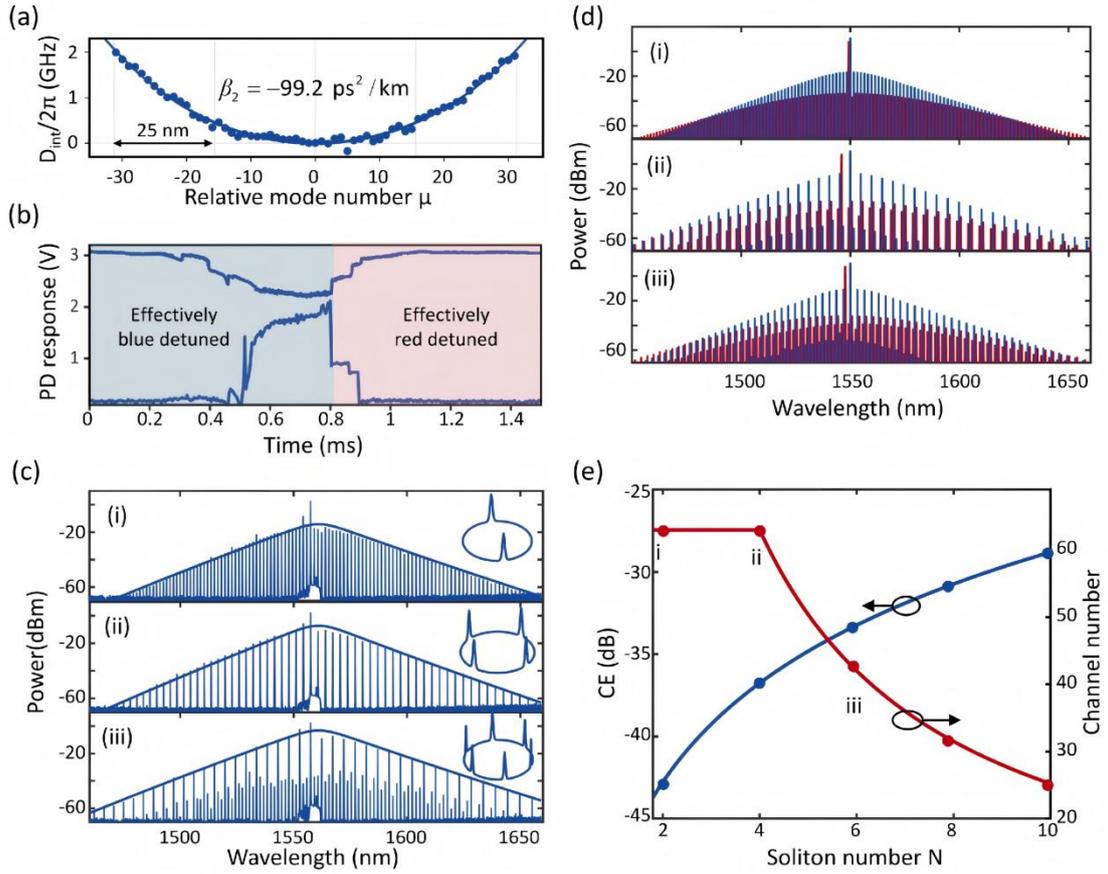

**Fig. 3 Experimental generation of perfect soliton crystals (PSCs) and simulations of signal broadcasting with PSCs.** (a) Measured dispersion of the high-Q resonances, giving a second-order dispersion coefficient of $\beta_2 = -99.2$ ps²/km. Dots and solid line correspond to measured integrated dispersion parameter $D_{int}(\mu)$ and the fitting curve, respectively. (b) Transmission (top) and comb teeth power (bottom) during the scan of the pump laser across a narrow resonance (no input signal light at this moment). (c) Measured optical spectra of 2-, 4-, 6-soliton crystal with smooth $sech^2$ envelopes. (d) Numerical simulation of broadcasting using different PCSs as pump light, where the red color represents the signal light and the broadcast lights. (e) Conversion efficiency and broadcast channel number versus soliton number.

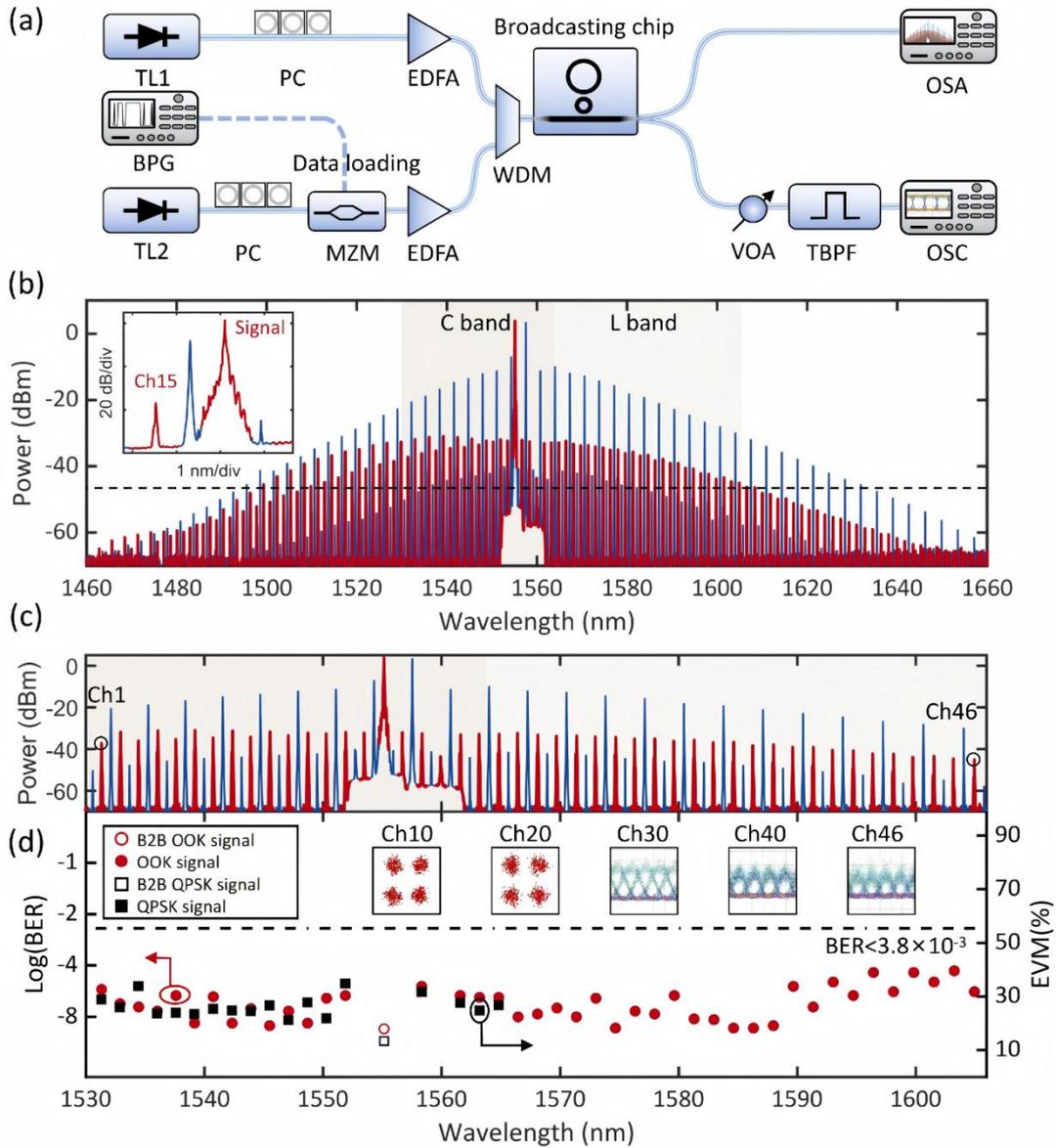

**Fig. 4 Soliton-assisted massive signal broadcasting.** (a) Schematic diagram of the soliton-assisted massive signal broadcasting setup. EDFA, erbium-doped fiber amplifier; PC, polarization controller; BPG, bit pattern generator; WDM, wavelength division multiplexer; TBPF, tunable bandpass filter; VOA, variable optical attenuator; OSA, optical spectrum analyzer; OSC, oscilloscope. (b) Optical broadcasting spectrum at the output of the bus waveguide (10% tapped light). (c) The zoomed-in optical spectrum of C and L band. (d) BER and EVM of the replicas and the original signals.

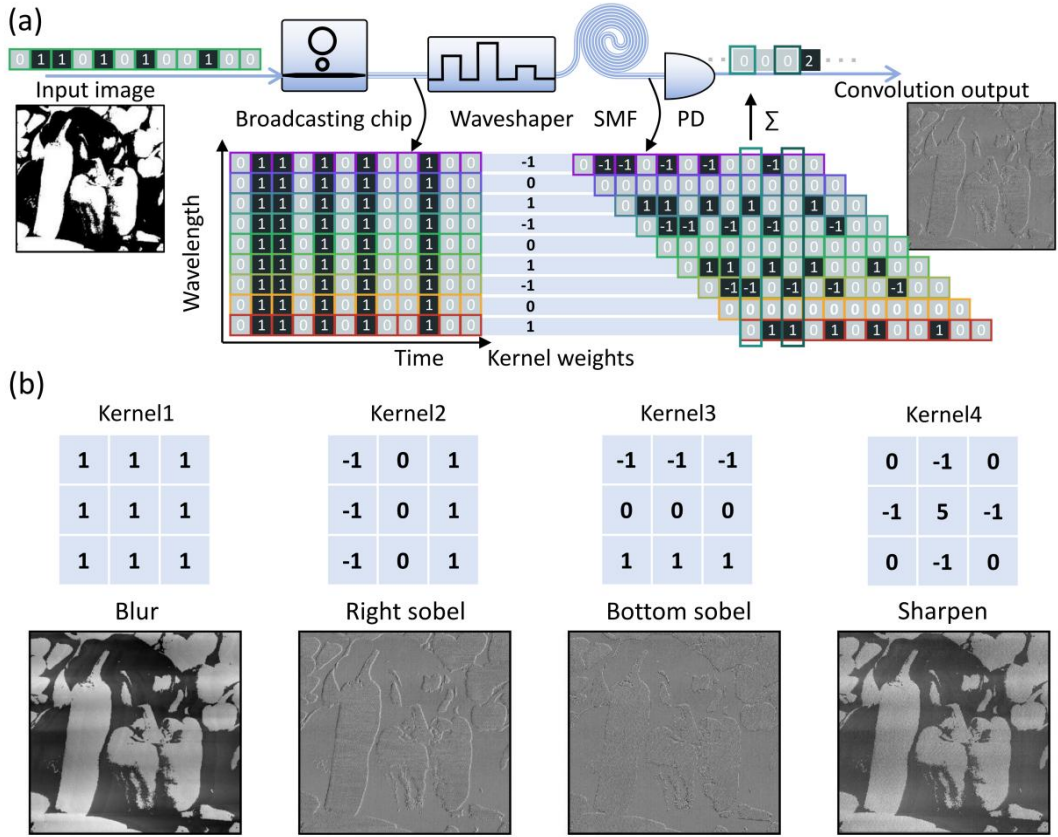

**Fig. 5 Massive signal broadcasting applied in optical convolutional accelerators (OCA).** (a) Experimental setup of image processing based on the proposed scheme with convolution kernel weights. SMF, single-mode fiber; PD, photodiode. (b) The weight matrices of the kernels and corresponding recovered images after OCA processing.

**Table 1 Comparison of on-chip all-optical signal broadcasting schemes**

| Material platform | Modulation Format | Baud rate (GBaud) | Channel count[a] | Aggregate data rate[b] (Tbps) | Broadcast range (nm) | Pump number[c] | Phase noise accumulation | Ref. |
|---|---|---|---|---|---|---|---|---|
| Si | DPSK | 10 | 28 | 0.15 | 25 | 2 | Yes | 5 |
| Si | OOK | 12 | 4 | 0.048 | 4 | 1 | No | 6 |
| PIN-Si | QPSK | 40 | 9 | 0.72 | 15 | 3 | Yes | 7 |
| PPLN | 16QAM | 10 | 8 | 0.32 | 11 | 9 | No | 8 |
| AlGaAs | PAM4 | 10 | 10 | 0.1 | 14 | 2 | Yes | 9 |
| InP | OOK | 40 | 2 | 0.08 | 25 | 1 | No | 10 |
| $Si_3N_4$ | QPSK | 15 | 100 | 1.26 | 200 | 1 | No | this work |

[a] The channel count considered here refers to the number of observable channels above the noise floor
[b] The aggregate data rate is calculated based on all measured channels with a BER below the required limit
[c] The pump number refers to the number of pump lasers used at the input

**Acknowledgements**

This work is supported by the Quantum Science and Technology-National Science and Technology Major Project (No. 2024ZD0300800); The National Natural Science Foundation of China (NSFC) (No. 62275087, 62475037, 62275090); The National Key Research and Development Program of China (No. 2023YFB2906305); The Hubei Provincial Natural Science Foundation of China (No. 2025AFA062, 2024AFB575); The Natural Science Foundation of Wuhan (No. 2025040601020213).

**Author contributions**

Z.F., Y.H. and W.D. contributed equally to this work. J.X, W.D. and H.Z. conceived the idea and designed the signal broadcasting and the OCA experiments. Z.F. and H.L. contributed to the coupled cavity device design. Z.F. and Y.H. carried out the experiments. W.D., H.Y., Y.C. and J.H. took part in various stages of the system building and data analysis. Z.F. and Y.H. processed the data and provided theoretical analysis and numerical simulation. J.X., Z.F., Y.H., W.D., N.C., H.Z., and X.Z. wrote the manuscript with contributions from all authors. J.X., H.Z. and X.Z. supervised the project.